\documentclass[11pt]{amsart}
\usepackage{geometry}                % See geometry.pdf to learn the layout options. There are lots.
\usepackage{graphicx}
\usepackage{amssymb}
\usepackage{epstopdf}
\usepackage{subfigure}

\title[Liquid Crystal Fingers Computers]{Computing with Liquid Crystal Fingers: \\ Models of geometric and logical computation} %inspired by localisations propagating \\ in liquid crystals}
\author[Adamatzky, Kitson, De Lacy Costello, Matranga, Younger]{Andrew Adamatzky$^1$,  Stephen Kitson$^2$,
Ben De Lacy Costello$^1$, Mario Ariosto Matranga$^2$ and Daniel Younger$^2$}
\begin{document}

\maketitle
\centerline{$^1$ University of the West of England, Bristol, BS16 1QY UK}
\centerline{$^2$  HP Labs, Bristol, BS34 8QZ  UK}

\begin{abstract}
When a voltage is applied across a thin layer of cholesteric liquid crystal, fingers of cholesteric alignment can form and propagate in the layer. In computer simulation, based
on experimental laboratory results, we demonstrate that these cholesteric fingers can solve selected problems of computational geometry, logic and arithmetics. We show that branching fingers approximate a planar Voronoi diagram, and
non-branching fingers produce a convex subdivision of concave polygons. We also provide a detailed blue-print and simulation of a one-bit half-adder functioning on the principles of collision-based computing, where the implementation is via collision of liquid crystal fingers with obstacles and other fingers.
\end{abstract}

\section{Introduction}

Liquid crystals (LCs) are fluids composed of anisotropic (usually rod-shaped) molecules that exhibit long range order. In the simplest phase, known as the nematic, the molecules tend to align in a common direction, called the director \emph{n}. In this paper we use a  cholesteric phase, in which the LC contains at least one component which is chiral so that the director adopts a helical structure through the LC. In a display, a thin layer of LC is enclosed between transparent substrates which contain electrodes to enable the application of a voltage.

LCs behave as elastic media and will adopt the director configuration that minimises the elastic energy of the system. In addition the molecules have anisotropic dielectric properties so that applying an electric field will tend to cause them to rotate. The configuration adopted by the LC director will therefore be determined by the combination of the boundary conditions imposed by the inside surfaces of the substrates, as well as the interaction between the electric field and the LC molecules. In the devices considered in this paper the inside surfaces of the substrates are designed to induce homeotropic alignment, that is with the director orthogonal to the substrates. The thickness of the LC layer is chosen to be close to the natural pitch of the cholesteric LC, which is the distance over which the director undergoes a full rotation. Under these conditions the natural tendency of the LC to twist is suppressed and the LC adopts a uniform untwisted configuration with the director orthogonal to the substrates. However, small perturbations to the system can upset this equilibrium and can cause the director to collapse into complex localised structures known as cholesteric fingers. These fingers consist of extended domains of twisted LC and a rich variety of structures and phases has been
reported~\cite{smalyukh_2005, pirkl_2001, ribiere_1994, baudry_1999}.

\begin{figure}[!tbp]
\centering
\includegraphics[width=0.49\textwidth]{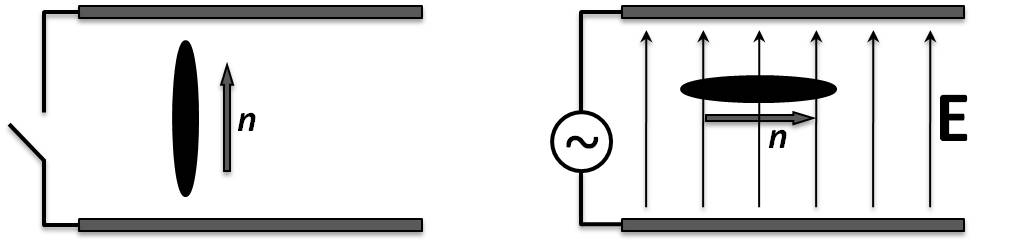}
\caption{The surfaces are treated so that with no voltage applied, the liquid crystal aligns with its director \emph{n} aligned normal to the substrates, as shown in the left figure. When the voltage is applied, the electric field \emph{E} rotates the LC, which has a negative dielectric anisotropy, so that it tends to align with the director parallel to the substrates.}
\label{lc}
\end{figure}

In our system we perturb the equilibrium by applying a voltage. The LC that we use has a negative dielectric anisotropy, that is the dielectric constant is greatest for electric fields directed orthogonal to the long axes of the rod shaped molecules. The field, which is applied across the layer, tends to rotate the director away from the initial vertical state
(Fig.~\ref{lc}), and as the director becomes more planar the natural chirality of the LC dominates resulting in the formation of cholesteric fingers. The anisotropic optical properties of the LC  means that these fingers are clearly visible under an optical microscope. In this paper we enhance the contrast by adding dichroic dyes~\cite{white_1974}. The fingers nucleate at defects or particles~\cite{smalyukh_2005} in the LC and grow as the applied voltage exceeds a threshold and retract as it reduces below the threshold. For larger voltages the fingers start to branch, and as they fill the space the fingers start to repel each other.

Fingers have also been reported in other LC phases~\cite{li_1995, wang_1998} and there have also been studies of isolated fingers which can propagate as a result of electroconvection in nematic LCs~\cite{dennin_1998, riecke_1998}. These propagating localisations in liquid crystal first attracted the attention of the unconventional computation community in early 2000s in the frame of collision-based computing. It was demonstrated in cellular automata models that worms in liquid crystals can implement basic logical gates when they collide with each other and either annihilate or deflect as a result of the collision~\cite{adamatzky_CBC} however no extended results were obtained at that time. Recent work in HP Labs has focussed on the interaction between LC materials and engineered microstructures~\cite{kitson_2002, kitson_2008} and some of these systems have shown the potential for engineering the nucleation and propagation of cholesteric fingers. These results inspired us to reconsider the concept of collision-based computing in liquid crystals and to expand a range of problems solved by the propagating and interacting fingers.

The paper is structured as follows.  In Sect.~\ref{experiment} we describe how we make our samples and measurements and in Sect.~\ref{model} we show how to imitate space-time dynamics of LC finger propagation using mobile automata on lattices. The classical problem of the planar Voronoi diagram is tackled by LC fingers in Sect.~\ref{voronoi}. Subdivision of a concave shape in convex shapes using LC fingers is demonstrated in Sect.~\ref{concave}.  We describe a design of a one-bit half-adder implemented via collision between fingers and obstacles, and between pairs of fingers, in Sect.~\ref{adder}.  Final thoughts of future LC finger computing devices are outlined in Sect.~\ref{discussion}.

\section{Fabrications and measurement of devices}
\label{experiment}

Devices were constructed using glass substrates coated with a transparent conductor layer made from ITO. In order to control the nucleation of the cholesteric fingers we patterned polymer structures onto one surface. Fingers tend to nucleate at the sharpest points on any structures. Fig.~\ref{experimental} shows a typical set of features: 40$\mu$m diameter rings with 4 protrusions to seed the nucleation of fingers. The structures are 5$\mu$m high and cover the ITO on one substrate. They are made from SU8 (Microchem Corp.) using photolithography. The surfaces of the two substrates are then coated with a thin polymer layer to align the LC homeotropically. To assemble the cell, the substrates are gently placed in contact so that the SU8 structures set the cell spacing, and the substrates are sealed on two edges with UV curing glue (NOA73 from Norland Products Inc.). The cell is then capillary filled with the LC mixture, with the cell heated to above the isotropic/nematic phase transition temperature to avoid the flow artifacts that can occur when filling with the LC in the nematic phase. The LC used was MLC2037 (Merck KGaA) which was chosen as it has a negative dielectric anisotropy, and the filling was carried out at $95^\circ$C. It was doped with an additive (1.29\% by weight of zli811, Merck) to impart a chiral pitch to the mixture. The concentration of the chiral dopant was chosen so that the pitch was very close to the cell spacing, so that the homeotropic state was just stable. The mixture was then doped with a blend of dichroic dyes (1\% by weight G232, G241 and G472 from Hayashibara Biochemical Laboratories, Inc.). These enhance the contrast and make the fingers visible, without the need to use polarisers~\cite{white_1974}. Dichroic dyes are rod-shaped molecules that only absorb light polarised along their long axis. They align with the LC, so that in the quiescent state the LC and dyes are both aligned vertically. Thus the dyes absorb very little light. When the fingers form, the LC and dyes at least partially align at a more planar angle, so that the dyes absorb more light and the fingers appear black.
Applying a voltage across the LC layer causes the fingers to nucleate from the sharp points of the structures. These then slowly extend while the voltage is maintained (Fig.~\ref{experimental}), and then retract when the voltage is removed. We use a 1kHz sinusoidal voltage with an amplitude of around 1.5V. In the next section we describe a model that captures some of the behaviour of the cholesteric fingers.
\section{Mobile automata model of liquid crystal fingers}
\label{model}

 A mobile automaton is a tuple $A = \langle c, s, f, \alpha, d, \rangle$, where
$c \in \mathbf{R}^2$ is an Euclidean coordinate of the automaton, $c = (x, y)$;
function $f: \mathbf{R}^2 \rightarrow \mathbf{R}^2$ transforms coordinates
as $x^{t+1}  = x^t + d \sin \alpha$ and $y^{t+1}  = y^t + d \cos \alpha$; in models
presented here $d=0.5$.
The automaton moves on a lattice $L$. All nodes of $L$ are in state 'empty' (0) initially. At each step $t$
of evolution time the nodes are updated as follows: for every $u \in L$: $u^t=1$ if
there is an automaton $A$ with coordinates $u$. The 'occupied' state is absorbing.
Obstacles are also defined as domains of 'occupied' states '1'.

\begin{figure}[!tbp]
\centering
\subfigure[]{\includegraphics[scale=0.3]{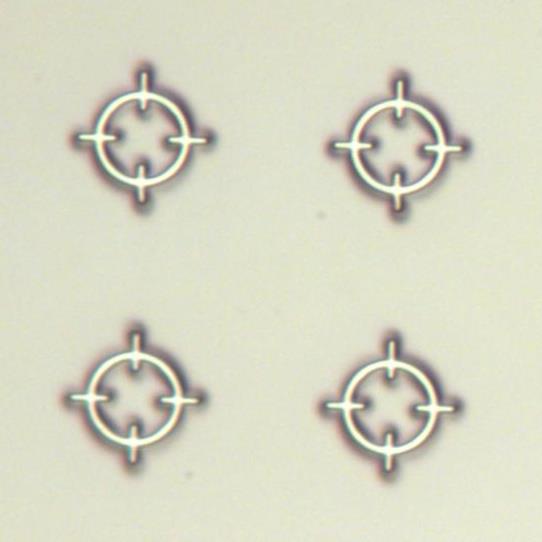}}
%\subfigure[]{\includegraphics[scale=0.3]{figs/experimentalcross/exp2}}
\subfigure[]{\includegraphics[scale=0.3]{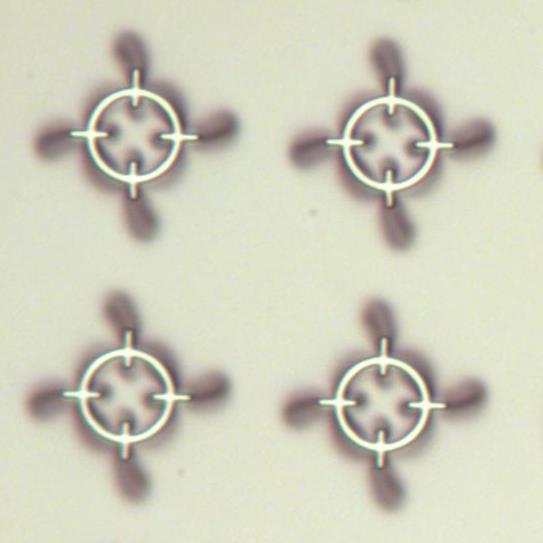}}
\subfigure[]{\includegraphics[scale=0.3]{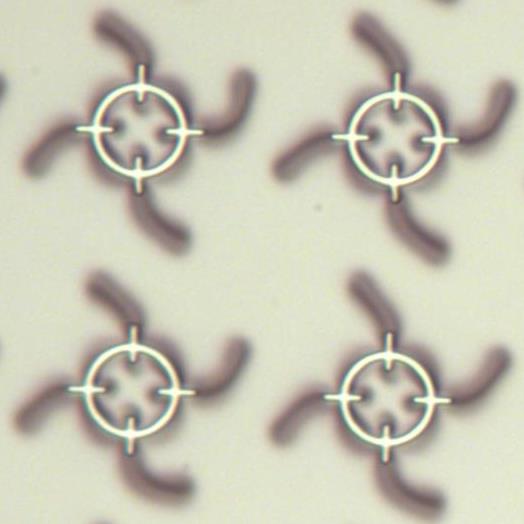}}
\subfigure[]{\includegraphics[scale=0.3]{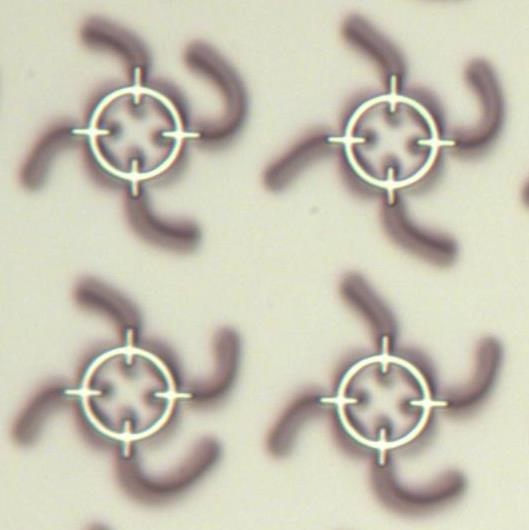}}
\subfigure[]{\includegraphics[scale=0.3]{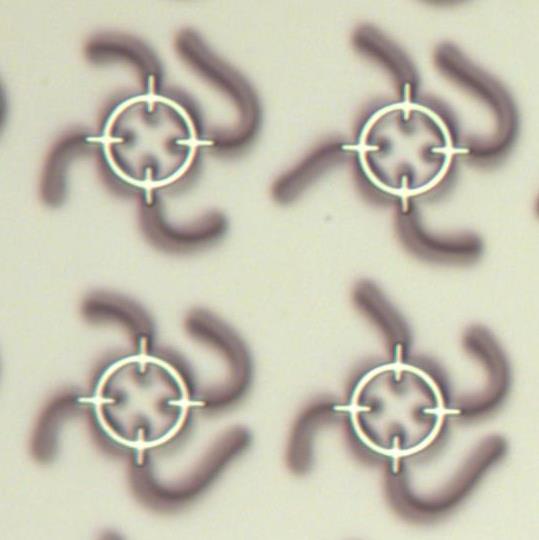}}
\subfigure[]{\includegraphics[scale=0.3]{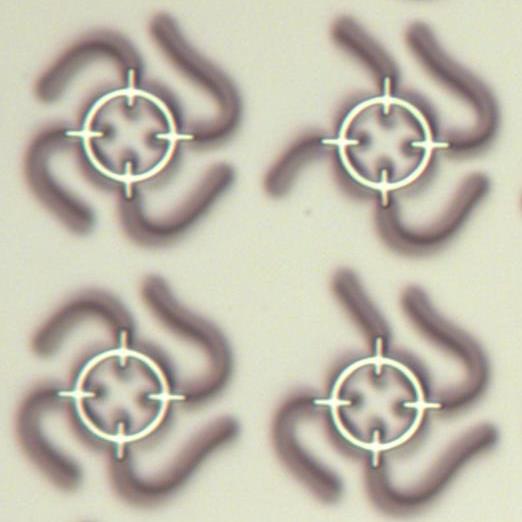}}
\caption{Experimental laboratory snapshots (optical microscopy) of colliding LC fingers, taken with an applied 1kHz ac voltage with an amplitude of 1.5V. Each photo is taken at a different time: (a) taken before the voltage is applied, (b) 4s, (c) 25s, (d) 40s, (e) 55s and (f) 80s after the voltage is applied}
\label{experimental}
\end{figure}

\begin{figure}[!tbp]
\centering
\subfigure[]{\includegraphics[scale=0.9]{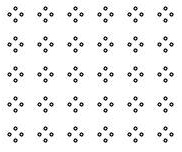}}
%\subfigure[]{\includegraphics[scale=0.7]{figs/smallcross/small_006}}
\subfigure[]{\includegraphics[scale=0.9]{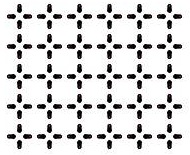}}
%\subfigure[]{\includegraphics[scale=0.7]{figs/smallcross/small_008}}
\subfigure[]{\includegraphics[scale=0.9]{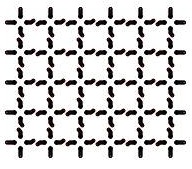}}
%\subfigure[]{\includegraphics[scale=0.7]{figs/smallcross/small_010}}
\subfigure[]{\includegraphics[scale=0.9]{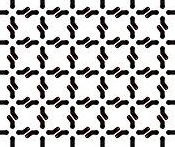}}
%\subfigure[]{\includegraphics[scale=0.7]{figs/smallcross/small_012}}
%\subfigure[]{\includegraphics[scale=0.7]{figs/smallcross/small_013}}
%\subfigure[]{\includegraphics[scale=0.7]{figs/smallcross/small_014}}
\subfigure[]{\includegraphics[scale=0.9]{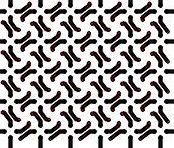}}
\subfigure[]{\includegraphics[scale=0.9]{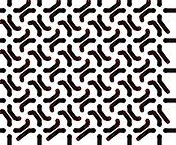}}
%\subfigure[]{\includegraphics[scale=0.7]{figs/smallcross/small_017}}
\caption{Simulation of experimental results, shown in Fig.~\ref{experimental},
using mobile automata. }
\label{simulated}
\end{figure}

\begin{figure}[!tbp]
\centering
\includegraphics[width=0.49\textwidth]{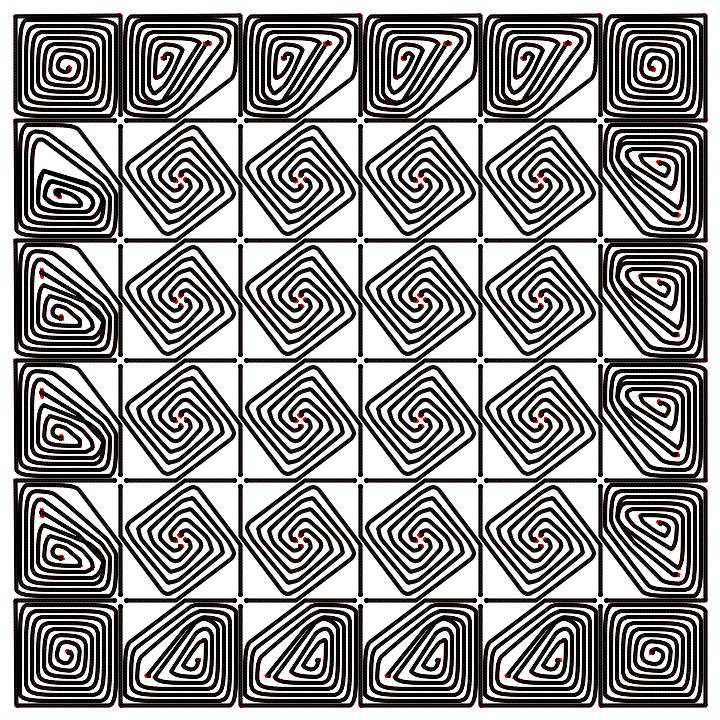}
\includegraphics[width=0.49\textwidth]{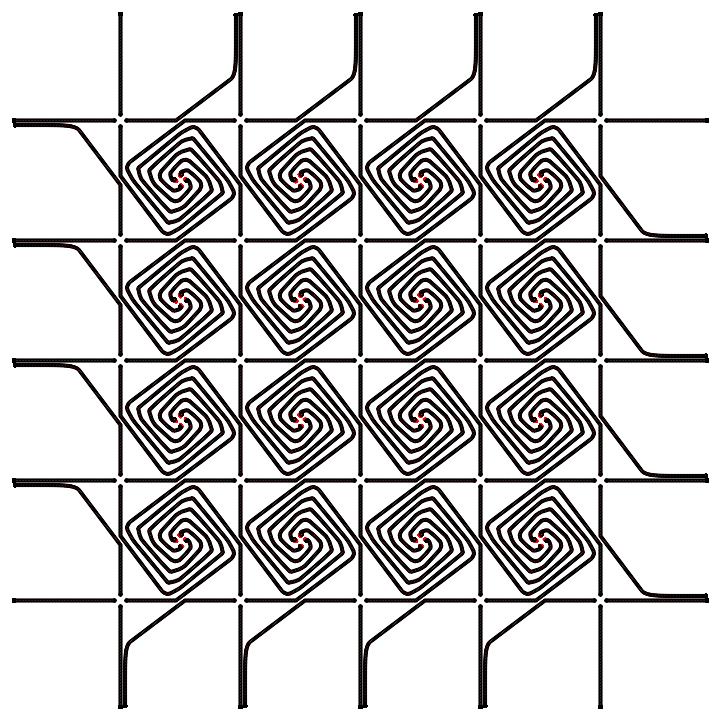}
\caption{Simulated interaction between fingers when distance between initial seeds is large enough to allow for
prolonged movement. (a)~Solid/impassable boundaries. (b)~Boundary-less space.}
\label{crosseslargerdistance}
\end{figure}

Fig.~\ref{experimental} shows a sequence of microscope photos of fingers propagating away from a dense array of nucleation sites. As fingers approach each other they deviate from a straight path to avoid contact. We find that the fingers always tend to turn in the same direction, left in the case of figure 1. We believe this is due to the natural chirality of the LC mixture. The model is based on these observations. An automaton $A$ chooses where to move as follows. If   node $(x^t + d \sin \alpha, y^t + d \cos \alpha)$ empty (0) then $A$ moves into this node; otherwise  it rotates left: $\alpha = \alpha +\pi/360$ and its state $s$ is incremented. The exact angle of scattering is not known and may depend on many factors beyond our present knowledge. Thus we adopted increment $\pi/360$ which $\alpha$ is increased with until a 'free' site for the next position is found. The automaton stops  when $s$ exceeds a  certain threshold of attempts.

The mobile automata model is phenomenological. A mobile automaton moving in a two-dimensional discrete space very roughly imitates the tip of a LC finger propagating in a quasi-two-dimensional space being squashed between two electrodes.
Our model in no way competes with existing numerical models of LC fingers,  see e.g. \cite{nagaya_1996}, but must rather be considered as a fast-prototyping tool in the design of LC finger computing algorithms.

We qualitatively verified our model using experimental observations. For example if we consider Fig.~\ref{experimental}.  Discs with four singularities/protrusions are arranged in a regular grid so that the protusions are aligned between neighbouring discs (the protusions of four neighbouring discs describe a square array) (Fig.~\ref{experimental}a). When a voltage of 1.5V is applied, LC fingers nucleate near the protrusions (Fig.~\ref{experimental}b). The fingers propagate along the original axis determined by the protrusions. When two fingers come into a head-on collision, each of them turns left  (Fig.~\ref{experimental}c--f). To imitate this experimental finding we regularly and uniformly distributed clusters of mobile automata in a two-dimensional space (Fig.~\ref{simulated}).  Initially the four automata of each cluster have their velocity vectors angles $\alpha$ oriented north, south, west and east (Fig.~\ref{simulated}a).  When finger $f'$ moving North collides with finger $f''$ moving south the finger $f'$ turns  south-east, and finger $f''$ north-west  (Fig.~\ref{simulated}b-d). Similarly, the  finger initially travelling west deviates  south-west and the finger travelling east turns north-east. Further changes of the fingers' trajectories are attributed to subsequent collisions  (Fig.~\ref{simulated}ef). If the distance between the initiation sites of the fingers is large enough to allow for prolonged movement of the fingers the four neighbouring fingers form spirals
(Fig.~\ref{crosseslargerdistance}).

\section{Approximation of Voronoi diagram}
\label{voronoi}

Let $\bf P$ be a non-empty finite set of planar points. A planar Voronoi diagram of the set $\bf P$ is a partition of the plane into such regions that, for any element of $\bf P$, a region corresponding to a unique point $p$ contains all those points of the plane which are closer to $p$ than to any other node of $\bf P$. A unique region $vor(p) = \{z \in {\bf R}^2: d(p,z) < d(p,m)\, \forall m \in {\bf R}^2, \, m \ne z \}$ assigned to the point $p$ is called a Voronoi cell of the point $p$. The boundary of the Voronoi cell of the   point $p$ is built up of segments of bisectors separating pairs of geographically closest points  of the given planar set $\bf P$. A union of all boundaries of the Voronoi cells determines the planar Voronoi diagram: $VD({\bf P}) = \cup _{p \in {\bf P}} \partial vor(p)$~\cite{shamos_preparata}.
Voronoi diagrams are applied in many fields of science and engineering. A few books and conference proceedings are available on the theory and applications of the Voronoi diagram~\cite{okabe_2000, anton_2009}.

Construction of a Voronoi diagram is a classical problem  of unconventional computing devices. This was the first ever problem solved in a reaction--diffusion chemical computer~\cite{tolmachev_adamatzky_1996}. The basic concept of constructing Voronoi diagrams with reaction--diffusion systems is based on an intuitive technique for detecting the bisector points separating two given points of the set $\bf P$. If we drop reagents at the two data points the diffusive waves,  or phase waves if the computing substrate is active, travel outwards from the drops. The waves travel the same distance from the sites of origin before they meet one another. The points where the waves meet are the bisector points; see the extensive bibliography in~\cite{adamatzky_2001,adamatzky_2005} and mechanisms of bisector formation in chemical media in~\cite{delacycostello_2003,costello_adamatzky_2003,ben_2004,ben_2004a,ben_2009}. The Voronoi diagram is also approximated in crystallisation-based processors~\cite{hotice} and colonies of acellular slime mould {\emph Physarum polycephalum}~\cite{AdamatzkyPhysarumMachines}. Thus it is intuitive to ascertain how Voronoi diagrams could be approximated with LC fingers.

\begin{figure}[!tbp]
\centering
\subfigure[]{\includegraphics[width=0.6\textwidth]{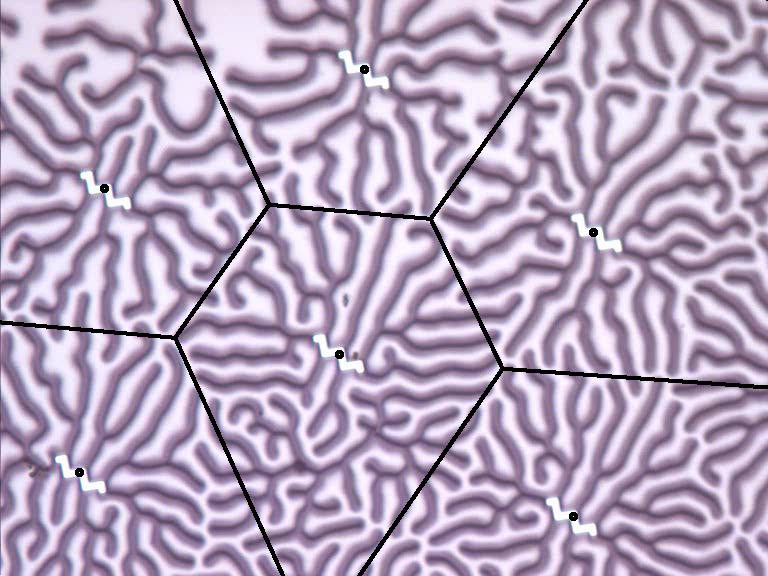}}
\subfigure[]{\includegraphics[width=0.6\textwidth]{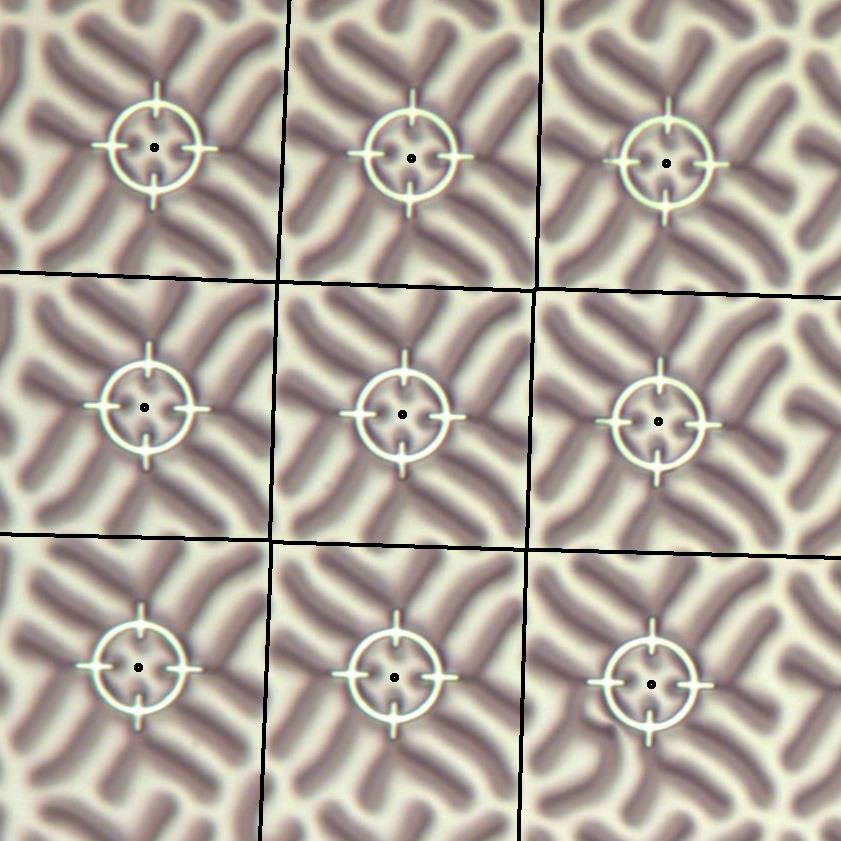}}
\caption{Experimental image of interaction between branching LC fingers originating from several seeds. The central point of the seeding structures are marked by small black discs. The edges of the Voronoi diagram are calculated using a
classical sweepline algorithm~\cite{fortune_1986} and are drawn as solid lines superimposed on the experimental image.}
\label{brachingexperimental}
\end{figure}

\begin{figure}[!tbp]
\centering
\subfigure[]{\includegraphics[width=0.32\textwidth]{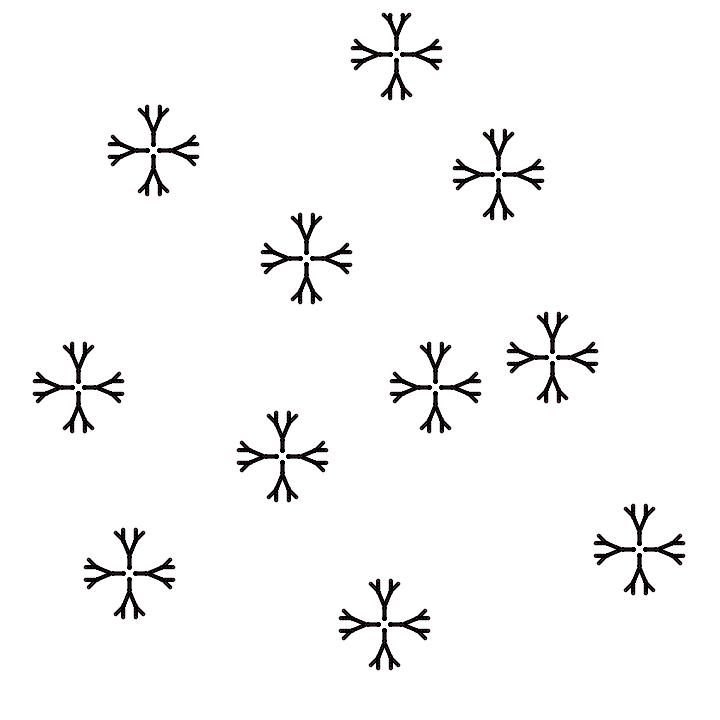}}
\subfigure[]{\includegraphics[width=0.32\textwidth]{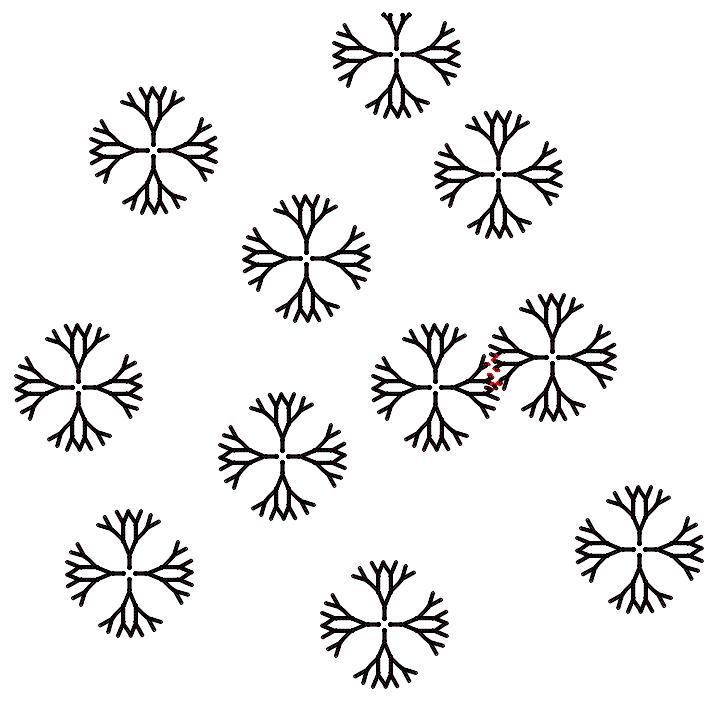}}
\subfigure[]{\includegraphics[width=0.32\textwidth]{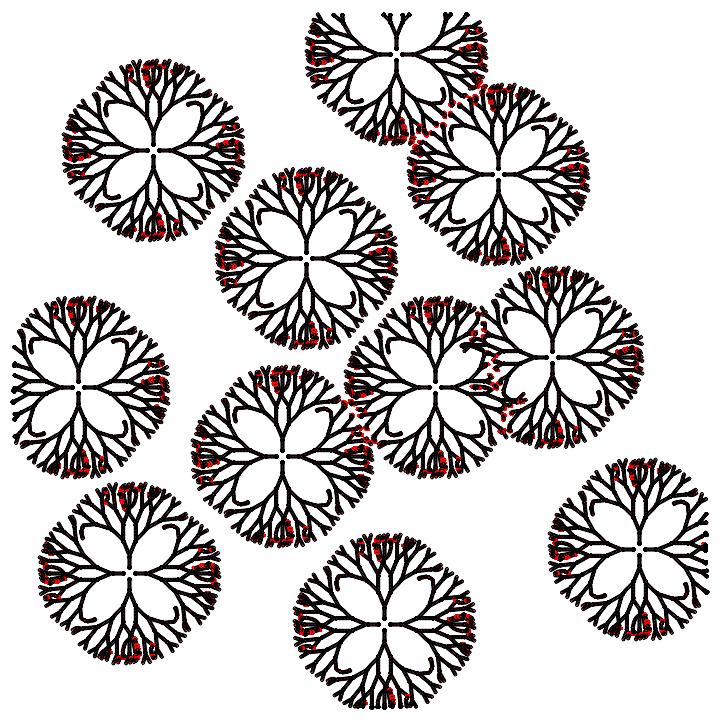}}
\subfigure[]{\includegraphics[width=0.32\textwidth]{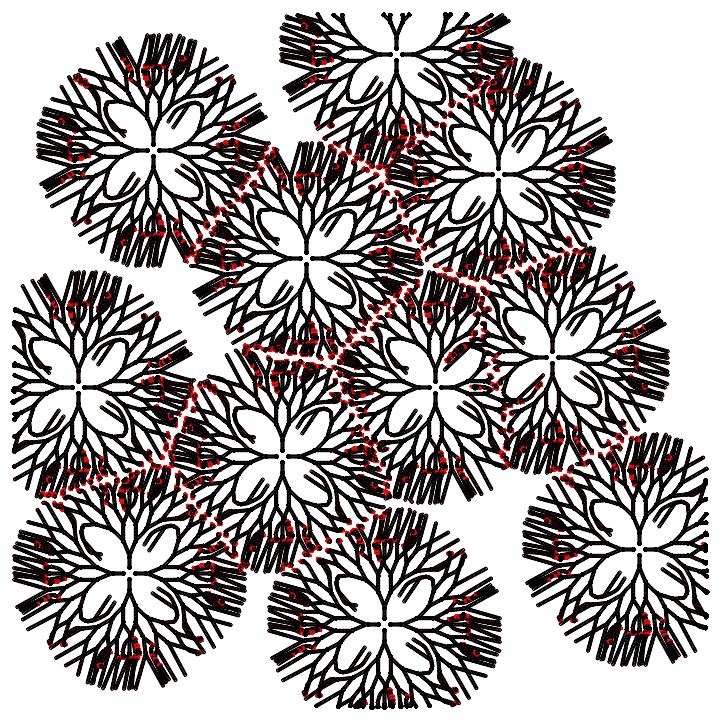}}
%\subfigure[]{\includegraphics[width=0.32\textwidth]{figs/branching/30114627}}
\subfigure[]{\includegraphics[width=0.32\textwidth]{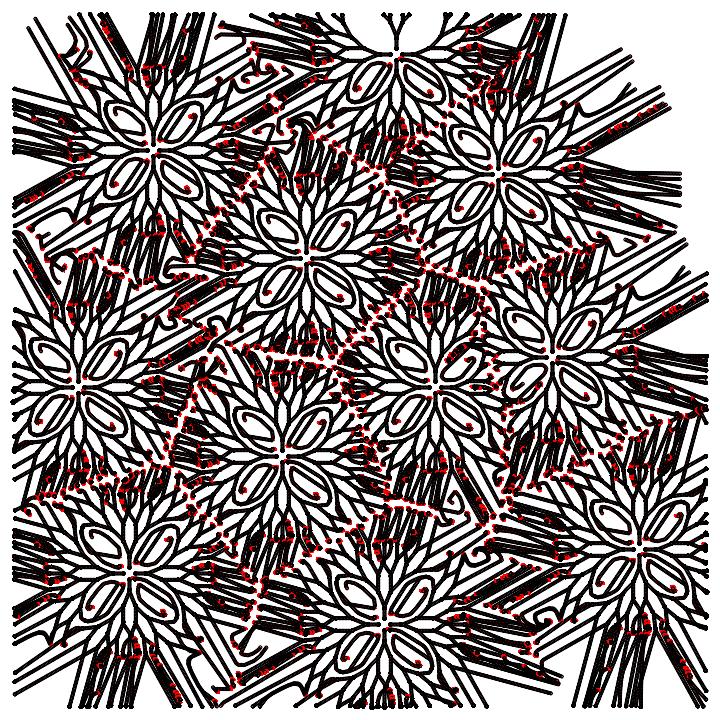}}
\subfigure[]{\includegraphics[width=0.32\textwidth]{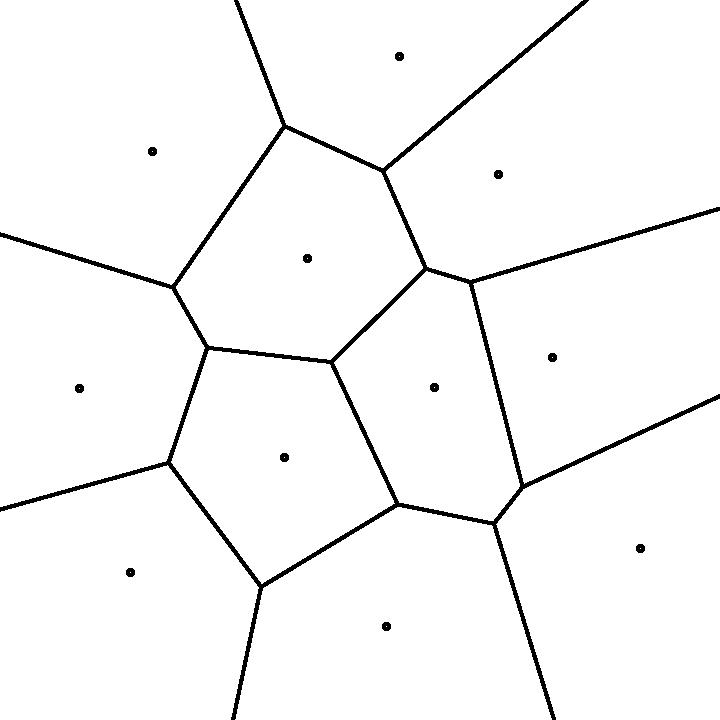}}
\caption{Simulated approximation of Voronoi diagram.
(a--e)~Snapshots of the simulated dynamics of branching LC fingers.
Edges of the  Voronoi diagram calculated using a classical sweepline algorithm~\cite{fortune_1986} are drawn by
solid red/gray lines in~(a--e) and shown explicitly in~(f).}
\label{branchingsimulated}
\end{figure}

The behaviour of the fingers is voltage-dependent. When the voltage exceeds a critical voltage (in this case 1.7 V)the fingers grow much more quickly and start to branch (Fig.~\ref{brachingexperimental}).  Recursively branching fingers form
fronts or a phase boundary between a free space and the domain filled by fingers. Fronts originating from different sources compete for space.  When two or more fronts meet, they stop propagating. Thus to approximate a Voronoi diagram we place the seeding structures of fingers in positions of space corresponding to planar points from $\bf P$ and increase the voltage to cause finger branching.The loci of space not occupied by any fingers represents the edge of the Voronoi diagram $VD({\bf P})$. From experimental observation it can be concluded that the accuracy of Voronoi diagram construction is improved where fingers are highly branched (approximate a continuous expanding front emanating from a defined site). This can be achieved by controlling the voltage or maximising the density of fingers initiated by altering the geometry and positioning of the seeding structures  (for example compare construction in
Figs.~\ref{brachingexperimental}a and ~\ref{brachingexperimental}b).
The approximation of the Voronoi diagram of eleven planar sites in mobile automata model of LC fingers is shown in Fig.~\ref{branchingsimulated}.

\section{Convex subdivision of concave polygons}
\label{concave}

\begin{figure}[!tbp]
\centering
\subfigure[]{\includegraphics[width=0.3\textwidth]{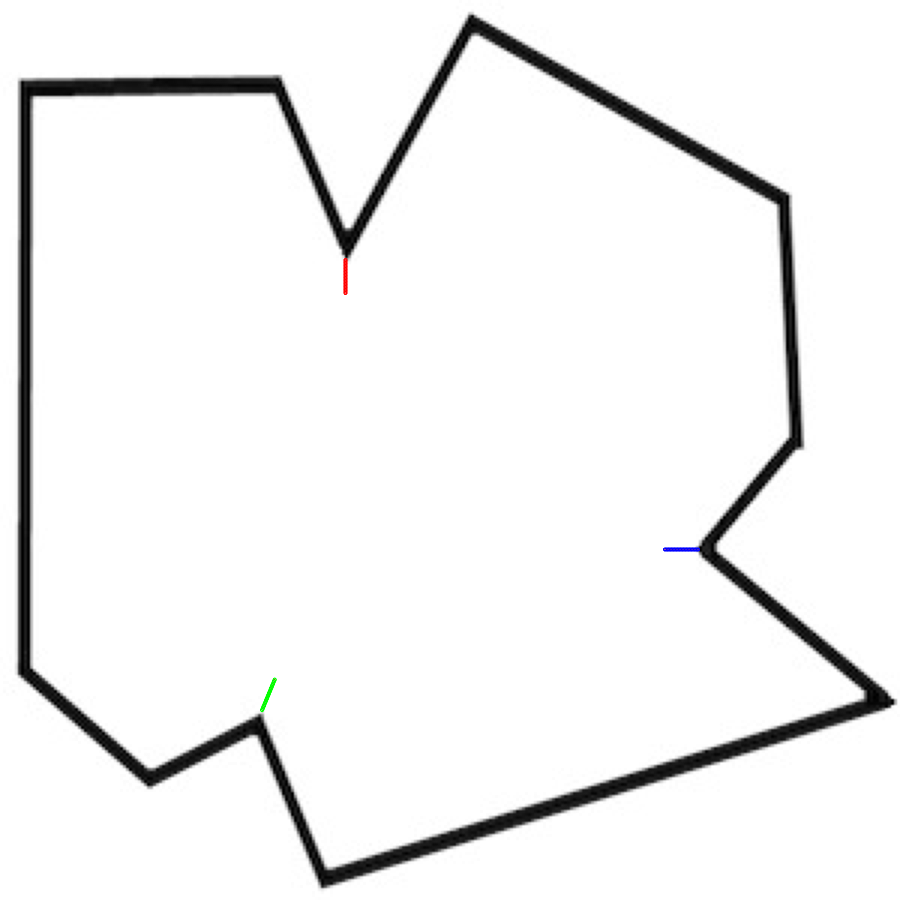}}
\subfigure[]{\includegraphics[width=0.3\textwidth]{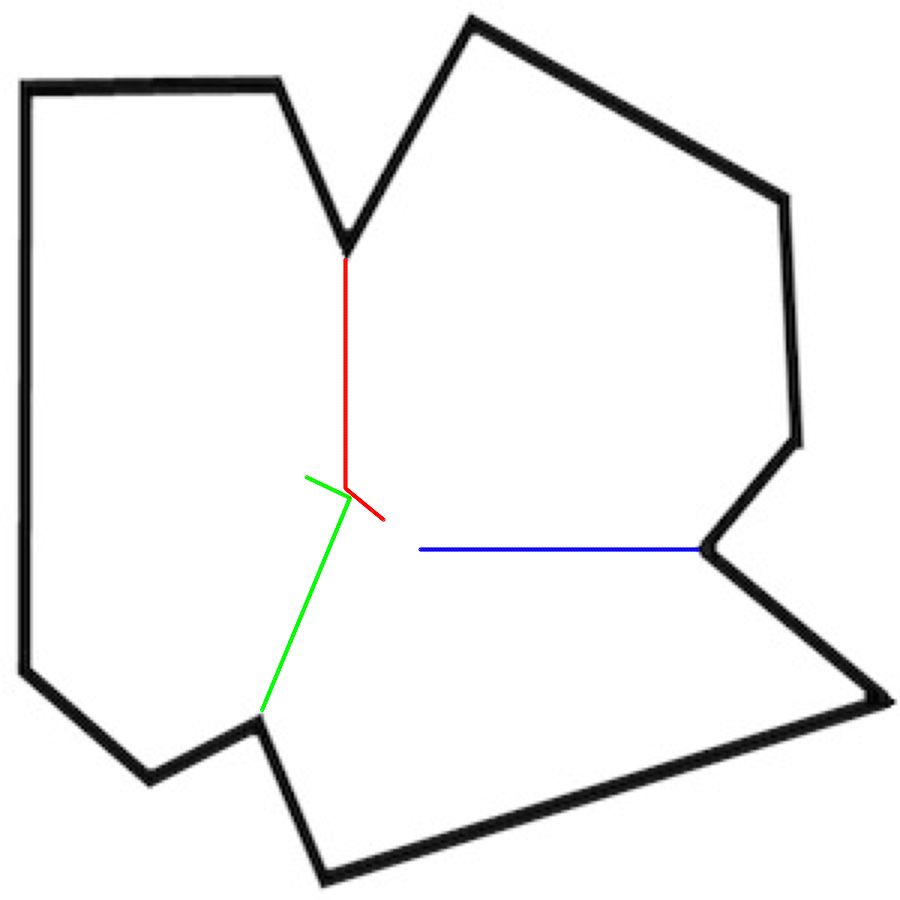}}
\subfigure[]{\includegraphics[width=0.3\textwidth]{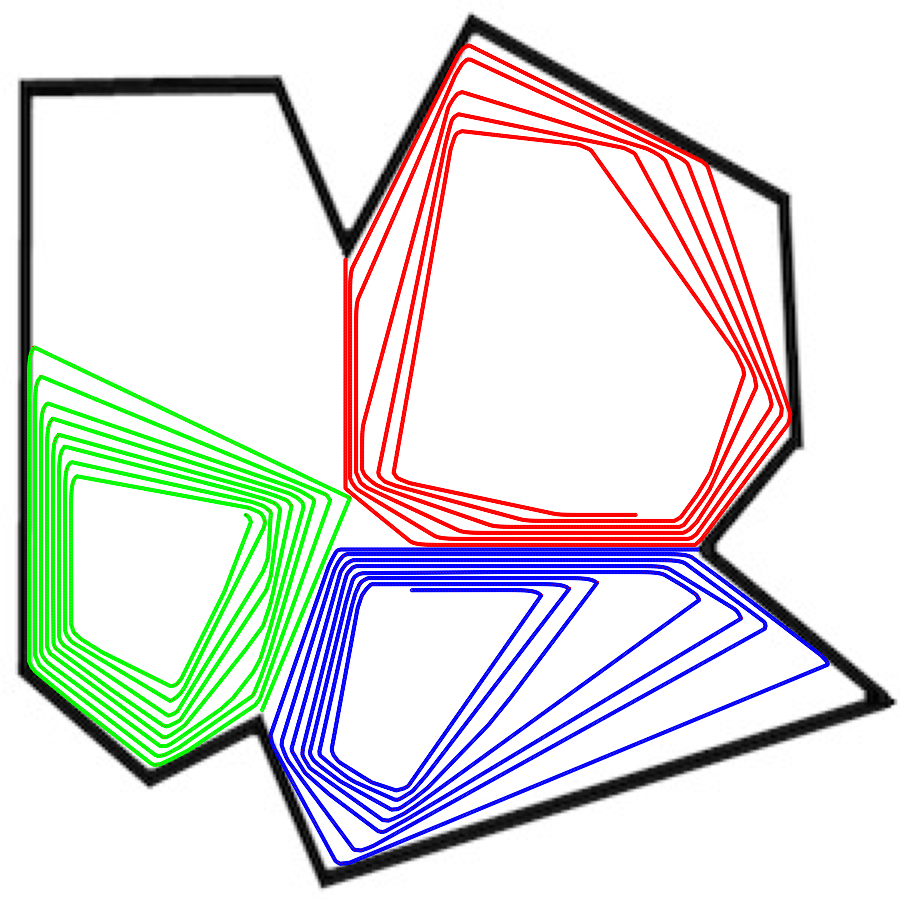}}
\subfigure[]{\includegraphics[width=0.3\textwidth]{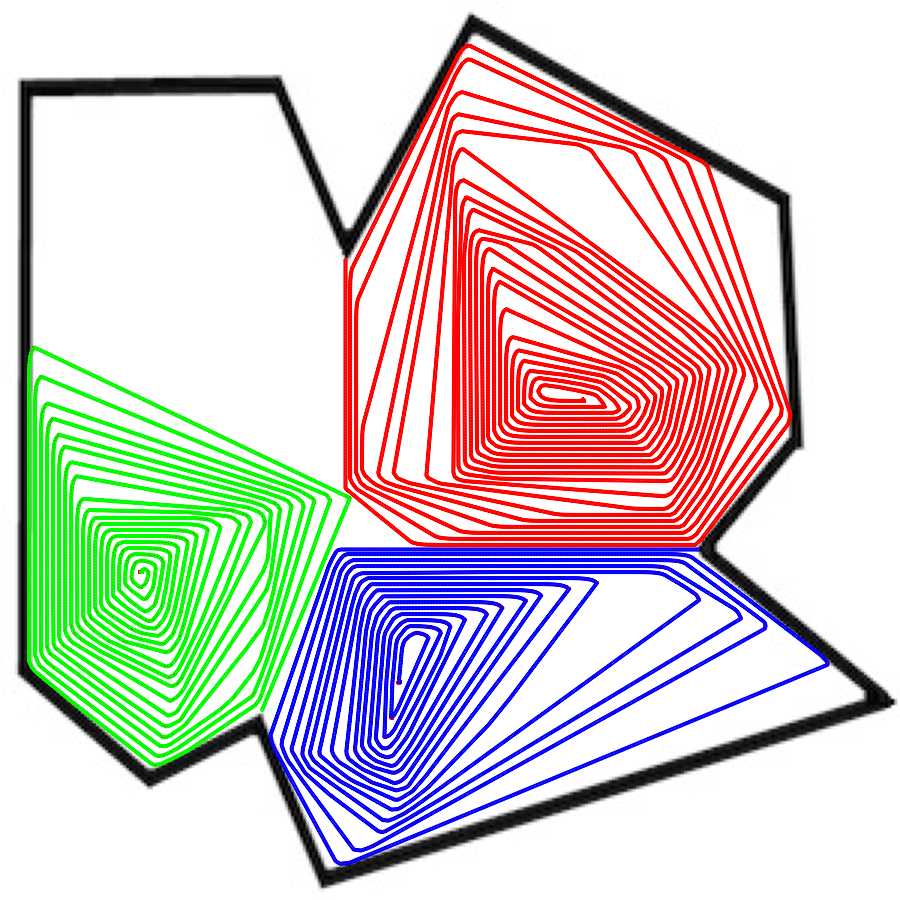}}
\subfigure[]{\includegraphics[width=0.3\textwidth]{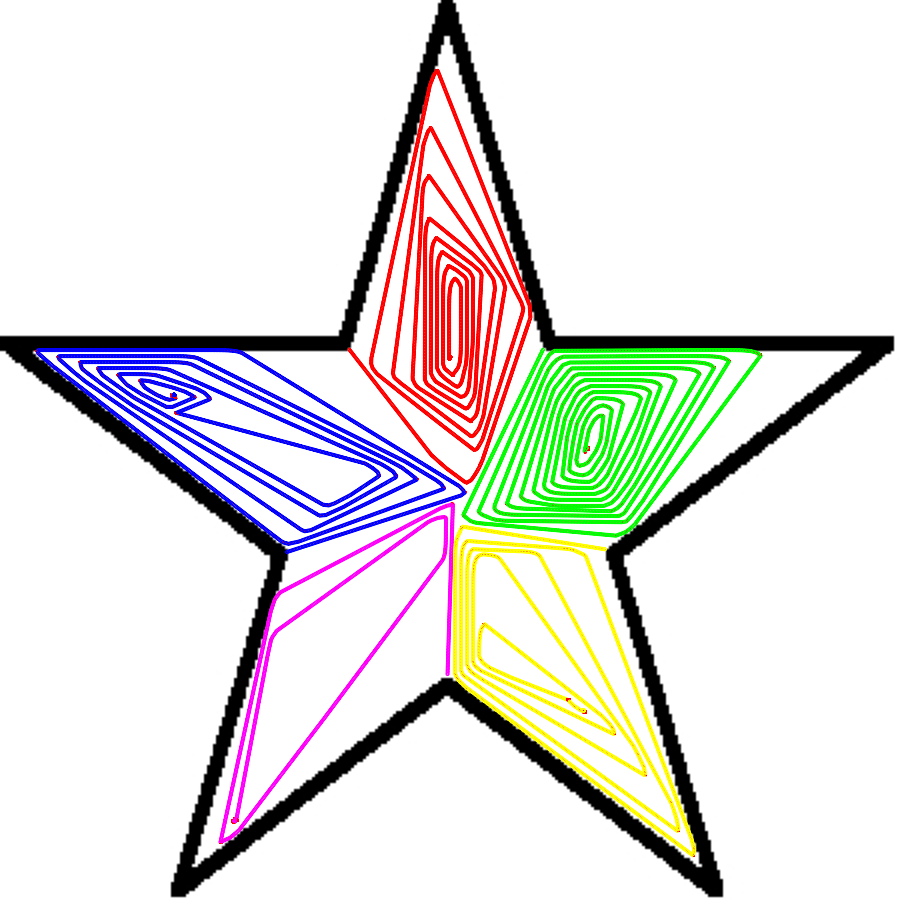}}
\caption{Convex subdivision of concave polygons in mobile automata model of LC fingers. Snapshots of the space after the propagation of
all fingers has stopped. (a--d)~Snapshots of LC finger model subdividing a concave polygon with three indentations. (e)~Convex subdivision of star-shaped polygon.}
\label{convexsubdivision}
\end{figure}

A Concave polygon is a shape comprised of straight lines with at least one indentation, or angle pointing inward.
The problem is to subdivide the given concave shape into convex shapes. The problem is solved by LC fingers as follows.
Fingers are generated at the singular points of indentations and the fingers' propagation vectors are co-aligned with medians of the corresponding inward angles. Given a concave polygon every indentation initiates
one propagating finger. A finger turns left, relative to its vector of propagation, when
it collides with another finger or a segment of the polygon. Given a polygon with $n$ indentations,
$n$ fingers will be generated. By following their ``turn-left-if-there-is-no-place-to-go''
routine and also competing for the available space with each other the fingers fill $n-1$ convex domains.
At least one convex domain will remain unfilled.

In the example shown in Fig.~\ref{convexsubdivision}a, there are  three indentations. They initiate fingers propagating --- approximately --- south ($a$), west ($b$) and north-north-east ($c$).  The finger travelling south ($a$)
collides with a finger travelling north-north-east ($c$). In the result of the collision, finger $a$ turn left and moves South-East. Finger $c$ turns left and moves North-West-West (Fig.~\ref{convexsubdivision}b). At some stage finger $b$ collides into finger $c$ and finger $a$ collides into finger $c$. When fingers collide to the 'walls' of the polygon they turn left and more or less accurately follow the wall till next collision (Fig.~\ref{convexsubdivision}c). Each finger forms a spiral and gets stuck inside the spiral at some point of its growth.

The computation of the convex subdivision of a concave polygon is completed when all fingers cease moving. The result of the computation is a set of convex polygons, each polygon is filled by a unique finger and at least one non-filled convex domain. In the example provided, a concave polygon is subdivided into four domains. The north-eastern domain is constructed by finger $a$, the south-eastern domain by finger $c$ and the south-western domain by finger $b$.  The north-western domain remains empty (Fig.~\ref{convexsubdivision}d). In the case of regular polygons, e.g. star-shaped in Fig.~\ref{convexsubdivision}e, boundaries between finger-filled convex domains correspond to the skeleton of this planar polygonal shape.

\section{One-bit half-adder}
\label{adder}

The half-adder presented is based on the paradigm of collision-based computing, which originates from
conservative logic and billiard ball model by Fredkin and Toffoli~\cite{fredkin_toffoli_1982} and logical computation
by colliding gliders in Conway's Game of Life~\cite{berlekamp_1992}. A collision-based computer employs mobile compact finite patterns and travelling localisations to represent quanta of information in non-linear media. Information values, e.g. truth values of logical variables, are given by either absence or presence of the localizations or other parameters of the localizations.  The localizations travel in space and when collisions occur the result can be interpreted as computation. There are no predetermined stationary signal channels (wires) --- however stationary localisations, or reflectors, are allowed --- a trajectory of the travelling pattern is considered a transient wire. Almost any part of the mediums' space can be used as a wire. Localizations can collide anywhere within the sample space, there are no fixed positions at which specific operations occur, nor location specified gates with fixed operations. The localizations undergo transformations, form bound states, annihilate or fuse when they interact with other mobile patterns. Information values of localizations are transformed as a result of collision~\cite{adamatzky_cbc}.

In the design discussed we use reflectors. The reflectors are stationary structures or defects. They are 'artificial'
in a sense that they do not belong to the liquid crystal medium but are externally and intentionally introduced to deflect
propagating fingers. We assume the obstacles are rectangular.  In real-world implementations
corners of the rectangular reflectors would generate additional propagating fingers, thus obstacles must be smooth e.g. circular or oval. However, our model is coarse-grained and therefore any oval reflector would be composed of small rectangles
and 'perceived' as a group of rectangles by the fingers.

\begin{figure}[!tbp]
\centering
\subfigure[]{\includegraphics[scale=0.3]{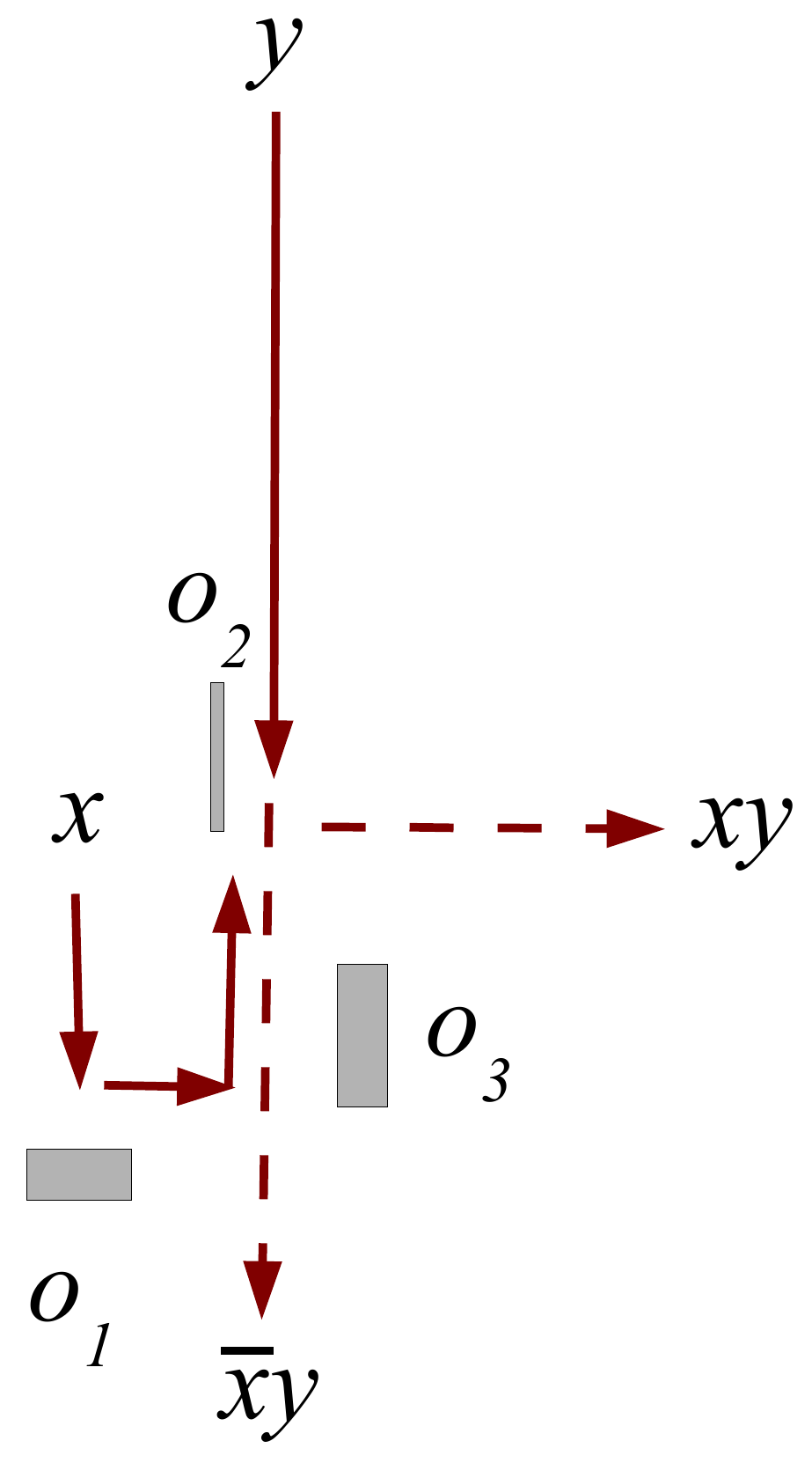}}
\subfigure[]{\includegraphics[scale=0.5]{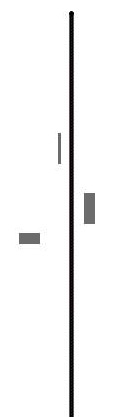}}
\subfigure[]{\includegraphics[scale=0.5]{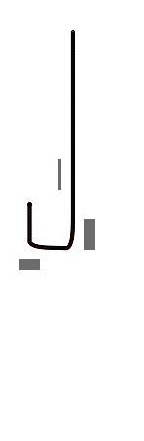}}
\subfigure[]{\includegraphics[scale=0.5]{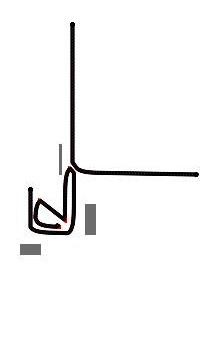}}
\caption{LC finger gate $\langle x, y \rangle \rightarrow \langle \overline{x}y, xy \rangle$.
(a)~Scheme: trajectories of fingers $x$ and $y$ entering gate are shown by solid lines, trajectories of fingers after
collision are shown by dotted lines, obstacles are shown by grey rectangles and tagged $o_1$, $o_2$ and $o_3$.
(b--d)~Configurations of two-dimensional mobile automata model.  Fingers and obstacles are represented by
states of cells.
(b)~Input $x=0$, $y=1$.
(c)~Input $x=1$, $y=0$.
(d)~Input $x=1$, $y=1$.}
\label{NOT(x)ANDy}
\end{figure}

Basic gate $\overline{x}y$ is illustrated in Fig.~\ref{NOT(x)ANDy}.  The gate consists of three obstacles $o_1$, $o_2$ and $o_3$ and two input fingers (Fig.~\ref{NOT(x)ANDy}a). The fingers represent values of Boolean variables $x$ and $y$: presence of a finger $x$ or $y$ corresponds to $x=1$ or $y=1$ ({\sc True}), absence of a finger corresponds to
0 ({\sc False}). Obstacles $o_1$ and $o_2$ are used to deflect finger $x$ while obstacle $o_3$ is for deflecting finger $x$ after collision with finger $y$.

Let us consider input $x=0$ and $y=1$ (Fig.~\ref{NOT(x)ANDy}b). Finger $x$ is not present. Finger $y$ travels south undisturbed. There are no obstacles in its way, thus it continues along its original trajectory which represents output
$\overline{x}y$. For input $x=1$ and $y=0$ only finger $x$ enters the gate (Fig.~\ref{NOT(x)ANDy}c). The finger $x$ turns east after colliding with obstacle $o_1$. It then collides with obstacles $o_3$ and turns north. Both outputs are nil.
When both inputs are  {\sc True} both fingers $x$ and $y$ enter the gate  (Fig.~\ref{NOT(x)ANDy}d). Finger $x$ enters the gate early than finger $y$. Thus by the time finger $y$ passes southward along obstacle $o_2$ finger $x$ moves northward. The fingers $x$ and $y$ collide with each other. In the result of this collision finger $y$ turns east and finger $x$ turns west.  The deflected trajectory of finger $y$ propagating east represents output $xy$. Deflected finger $x$ collides with obstacle $o_2$ deflected south, and eventually becomes trapped and propagation ceases (Fig.~\ref{NOT(x)ANDy}d).

\begin{figure}[!tbp]
\centering
\subfigure[]{\includegraphics[width=0.4\textwidth]{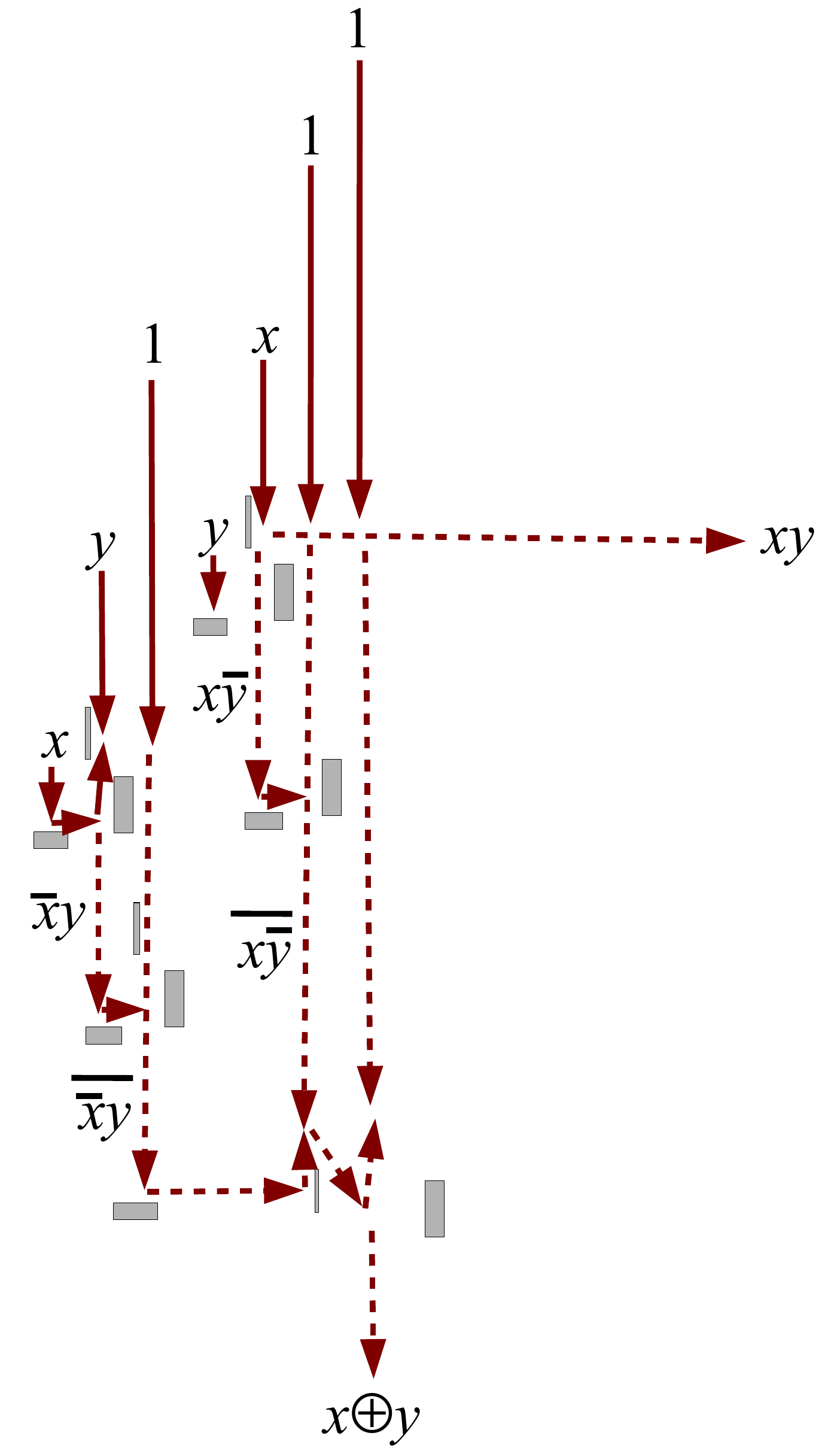}}
\subfigure[]{\includegraphics[width=0.3\textwidth]{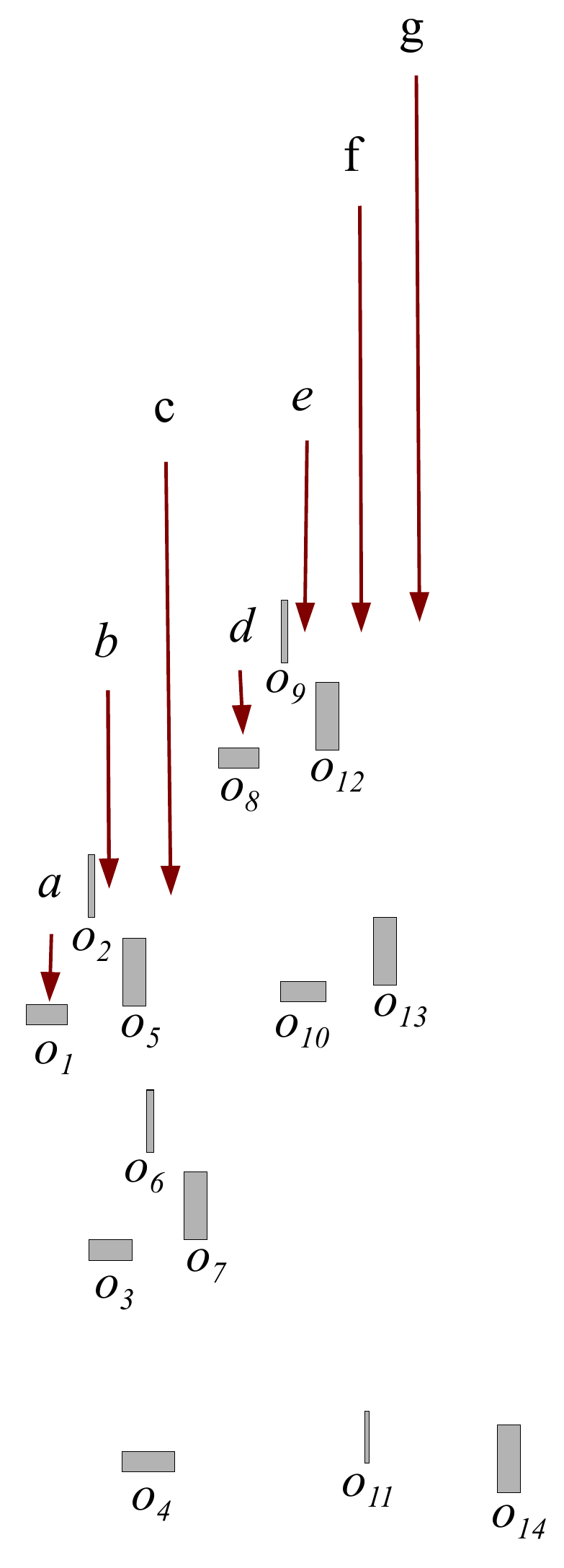}}
\caption{A scheme of one-bit half-adder implementable with LC fingers~(a) and labels of key elements of the adder~(b). }
\label{halfadderscheme}
\end{figure}

To implement a binary half-adder one must realise calculation of a sum and carry on results: $x \oplus y$ and $xy$. An architecture of LC finger half-adder can be implemented of three gates $\langle x, y \rangle \rightarrow \langle \overline{x}y, xy \rangle$ and five additional obstacles (Fig.~\ref{halfadderscheme}). There are seven input trajectories and two output trajectories. Inputs include three fingers representing constant {\sc Truth}, two fingers representing values of $x$ and two fingers representing values of $y$. In the present model we do not tackle multiplication or splitting of signals therefore we assume copies of variables $x$ and $y$ are presented \emph{a priori}. All input fingers travel south, output $x \oplus y$ travels south and output $xy$ travels east.  Note that all input fingers enter the half-adder at different moments of time (Fig.~\ref{halfadderscheme}a). In total we have seven fingers labelled $a$ to $g$ and fourteen obstacles labelled $o_1$ to $o_{14}$ (Fig.~\ref{halfadderscheme}b).

\begin{figure}[!tbp]
\centering
\subfigure[]{\includegraphics[width=0.33\textwidth]{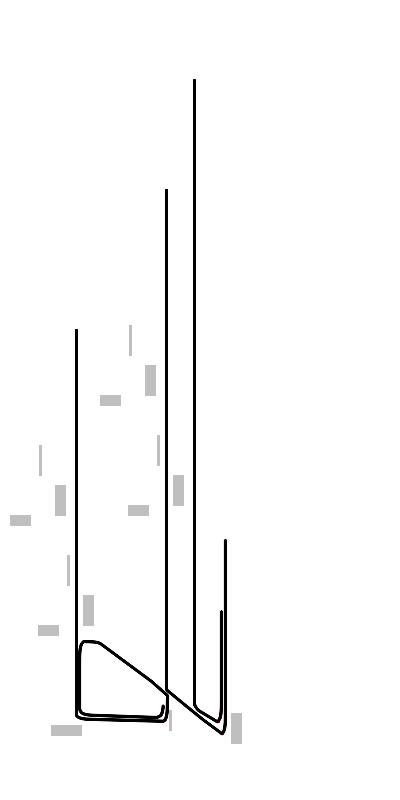}}
\subfigure[]{\includegraphics[width=0.33\textwidth]{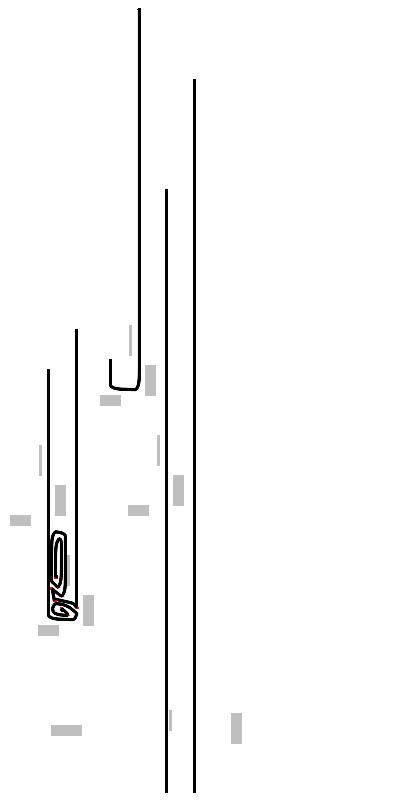}}
\subfigure[]{\includegraphics[width=0.33\textwidth]{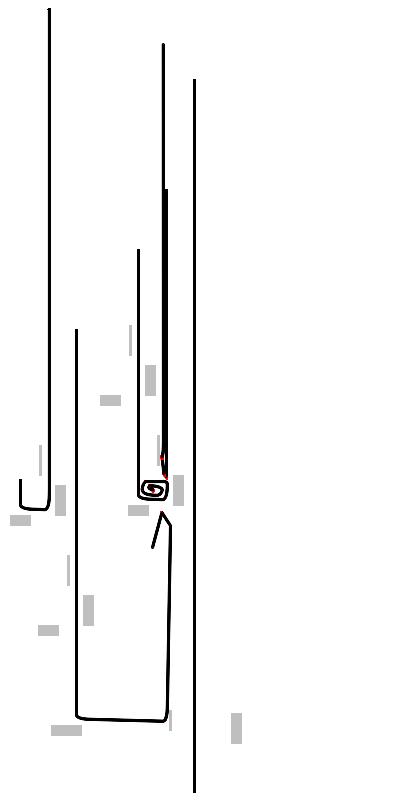}}
\subfigure[]{\includegraphics[width=0.33\textwidth]{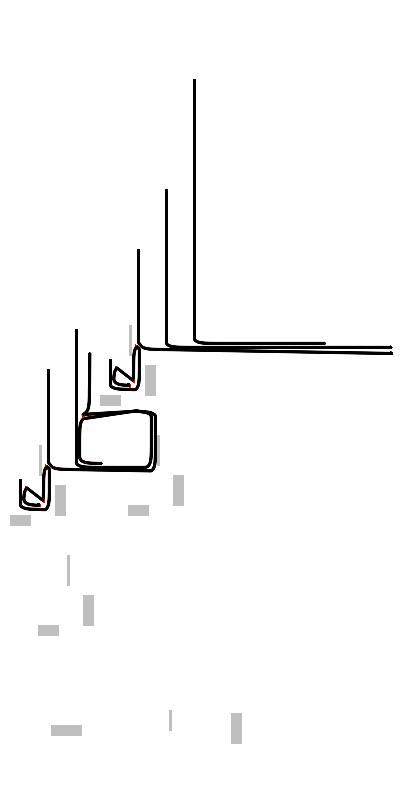}}
\caption{Configurations of two-dimensional mobile automata model.  Fingers and obstacles are represented by
states of cells.
(a)~$x=0$, $y=0$,
(b)~$x=0$, $y=1$,
(c)~$x=1$, $y=0$,
(d)~$x=1$, $y=1$.}
\label{halfadder}
\end{figure}

Simulation of one-bit half-adder in mobile automata model is shown is illustrated in Fig.~\ref{halfadder}.
When both inputs $x$ and $y$ are {\sc False} only three fingers corresponding to constant {\sc True} propagate southward (Fig.~\ref{halfadder}a). Finger $c$ collides is deflected by obstacle $o_4$, turns east, is deflected by
obstacle $o_{11}$ and turns north. It collides with finger $f$. In the result of this collision finger $c$ turns north-west and is "self-trapped" while finger $f$ turns south-east. The body of finger $f$ prevents further propagation of finger $g$. Finger $f$ and $g$ are deflected by obstacle $o_{14}$ and propagate north. Thus for input $x=0$ and $y=0$ no output fingers appear along dedicated trajectories  (Fig.~\ref{halfadder}a): $\langle 0, 0 \rangle \rightarrow \langle 0, 0 \rangle$.

For the input combination $x=0$ and $y=1$ the situation develops as follows (Fig.~\ref{halfadder}b). Fingers $c$, $f$ and $g$, representing constant {\sc True} enter the the adder as usual. Fingers $a$ and $e$, representing $x$, are absent. Fingers $b$ and $d$, representing $y=1$ enter the adder. Finger $b$ collides with obstacle $o_3$ and turns east, then collides with obstacle  $o_7$ and heads north. Thus fingers $b$ and $c$ come into  a head-on collision. Finger $b$ turns  west and is "self-trapped". Finger $c$ turns east, collides with obstacle $o_7$ and then $o_6$ and gets trapped. Finger $d$ collides with obstacle $o_8$, heads east, collides with obstacle $o_{12}$ and travels north.  Fingers $f$ and $g$ continue their travel undisturbed. Thus operation $\langle 0, 1 \rangle \rightarrow \langle 1, 0 \rangle$ is imlpemented.

When $x=1$ and $y=1$  (Fig.~\ref{halfadder}c) finger $a$ (representing first copy of $x$) is turned north by obstacles
$o_1$ and $o_5$. Finger $c$ (representing second copy of $x$) is turned north by obstacles $o_{10}$ and $o_{13}$,
collides with finger $f$. Fingers $c$ and $f$ stop propagating in result of the collision. Finger $c$ is turned east by obstacle $o_4$ and then north by obstacle $o_{11}$. Finger $g$ continues its propagation undisturbed and thus operation $\langle 1,  0 \rangle \rightarrow \langle 1, 0 \rangle$.

Most interactions between fingers take place in situation $x=1$ and $y=1$ (Fig.~\ref{halfadder}d). Finger $a$ collides with finger $b$. Finger $b$ turns east as a result of the collision. Body of finger $b$ prevent finger $c$ propagating north.
Finger $d$ collides with finger $e$ and gets "self-trapped". Finger $e$ turns east in result of the collision and exits the
adder. Final trajectory of finger $e$  represents $xy=1$ (Figs.~\ref{halfadderscheme}a and~\ref{halfadder}d).
Propagation  finger $g$ is blocked by body of fingers $e$ and $f$, it does not exit the adder along its original trajectory, and thus $x \oplus y = 0$.

\section{Discussion}
\label{discussion}

Cholesteric liquid crystals exhibit growth of localised phase defects --- fingers --- in a response to application of an ac electric field. The fingers show (almost) deterministic behaviour when they collide with other fingers and obstacles, thus they are suitable for implementation of collision-based computing schemes. For higher values of voltage applied fingers show branching. Wave-fronts of branching fingers stop their propagation when they collide with other fronts. Thus branching fingers can approximate Voronoi diagram, a plane subdivision based on proximity criteria. For low applied voltages the fingers remain solitary and thus each finger can represent a quantum of information and be an elementary unit of a collision-based computing device. To illustrate feasibility of the approach we provided design of a one-bit binary half-adder and proved correctness of its functioning using a mobile automata model.  Collision-based computing prototypes presented in the paper are
based on computer imitations of finger propagation and further work is required to implement the design in physical laboratory conditions.

\end{document}